\documentclass{article}
\usepackage{spconf,amsmath,graphicx}
\usepackage{booktabs}
\usepackage{color,amsmath,url,times, tabularx,bbm,amssymb,multirow}
\usepackage{hyperref}

\title{NoreSpeech: Knowledge Distillation based Conditional Diffusion Model for Noise-robust Expressive TTS}

\name{Dongchao Yang$^{1}$, Songxiang Liu $^{2}$, Jianwei Yu $^{2}$, Helin Wang$^{3}$, Chao Weng $^{2}$,
      Yuexian Zou$^{1*}$\thanks{Work done during an internship at Tencent AI Lab.}\thanks{This paper was partially supported by Shenzhen Science \& Technology Research Program (No:GXWD20201231165807007-20200814115301001) and NSFC (No: 62176008).}}
\address{$^{1}$ADSPLAB, School of ECE, Peking University, China\\
        $^{2}$ Tencent AI Lab, Shenzhen, China, $^{3}$ Johns Hopkins University, USA\\
}
\begin{document}
%
\maketitle
\begin{abstract}
Expressive text-to-speech (TTS) can synthesize a new speaking style by imiating prosody and timbre from a reference audio, which faces the following challenges: (1) The highly dynamic prosody information in the reference audio is difficult to extract, especially, when the reference audio contains background noise. (2) The TTS systems should have good generalization for unseen speaking styles. In this paper, we present a \textbf{no}ise-\textbf{r}obust \textbf{e}xpressive TTS model (NoreSpeech), which can robustly transfer speaking style in a noisy reference utterance to synthesized speech. Specifically, our NoreSpeech includes several components: (1) a novel DiffStyle module, which leverages powerful probabilistic denoising diffusion models to learn noise-agnostic speaking style features from a teacher model by knowledge distillation; (2) a VQ-VAE block, which maps the style features into a controllable quantized latent space for improving the generalization of style transfer; and (3) a straight-forward but effective parameter-free text-style alignment module, which enables NoreSpeech to transfer style to a textual input from a length-mismatched reference utterance. Experiments demonstrate that NoreSpeech is more effective than previous expressive TTS models in noise environments. Audio samples and code are available at: \href{http://dongchaoyang.top/NoreSpeech\_demo/}{http://dongchaoyang.top/NoreSpeech\_demo/}
\end{abstract}
\section{Introduction}
\label{sec:intro}
Text-to-speech (TTS) \cite{shen2018natural,ren2020fastspeech} aims to transform text into almost human-like speech, which attracts broad interest in the deep learning community. Nowadays, TTS models have been extended to more complex scenarios, including multiple speakers timbre, emotions, and speaking styles for expressive and diverse voice synthesis \cite{huang2022generspeech}. 
Style modeling and transferring have been studied for decades in the TTS community: Wang \textit{et al.} \cite{wang2018style} proposed to use global style tokens to control and transfer the global style. Li \textit{et al.} \cite{li2021towards} adopt a multi-scale style encoder to assist synthesis expressive speech. Min \textit{et al.} \cite{min2021meta} proposed Meta-StyleSpeech, which uses a meta-learning training strategy for multi-speaker TTS synthesis. Huang \textit{et al.} \cite{huang2022generspeech} proposed a multi-level style adaptor to transfer speaking style. However, these methods assume that the reference audio is recorded in ideal environments (without noise interference). This assumption prevents expressive TTS models from being applied in many real-world scenarios, \textit{e.g.} the reference audio recorded by users may include noise. Zhang \textit{et al.} \cite{zhang2022hifidenoise} proved that the fundamental frequency (F0) and energy can be affected by adding noise, which are key components of speaking style.
To eliminate the effect of noise in reference audio, many methods have been proposed \cite{goswami2022satts,lee2021styler,hsu2019disentangling,nikitaras2022fine}. These methods can be classified into two types: (1) using a pre-trained speech enhancement model to eliminate noise in reference audio \cite{goswami2022satts}, which heavily relies on the performance of a speech enhancement (SE) model; (2) decomposing the noise information via adversarial training \cite{hsu2019disentangling} or information bottleneck \cite{lee2021styler, nikitaras2022fine}. However, the adversarial training and information bottleneck strategies need complex parameter setting and training tricks, which makes them hard to be widely applied. In summary, all of these methods try to directly separate noise information from noisy reference and then extract the style information from the remaining parts. However, they ignore the diversity of noise and the highly dynamic time-frequency information in noise is hard to remove. 
\begin{figure*}[t] \label{norespeech}
  \centering
  \includegraphics[width=\linewidth, height=0.36\linewidth]{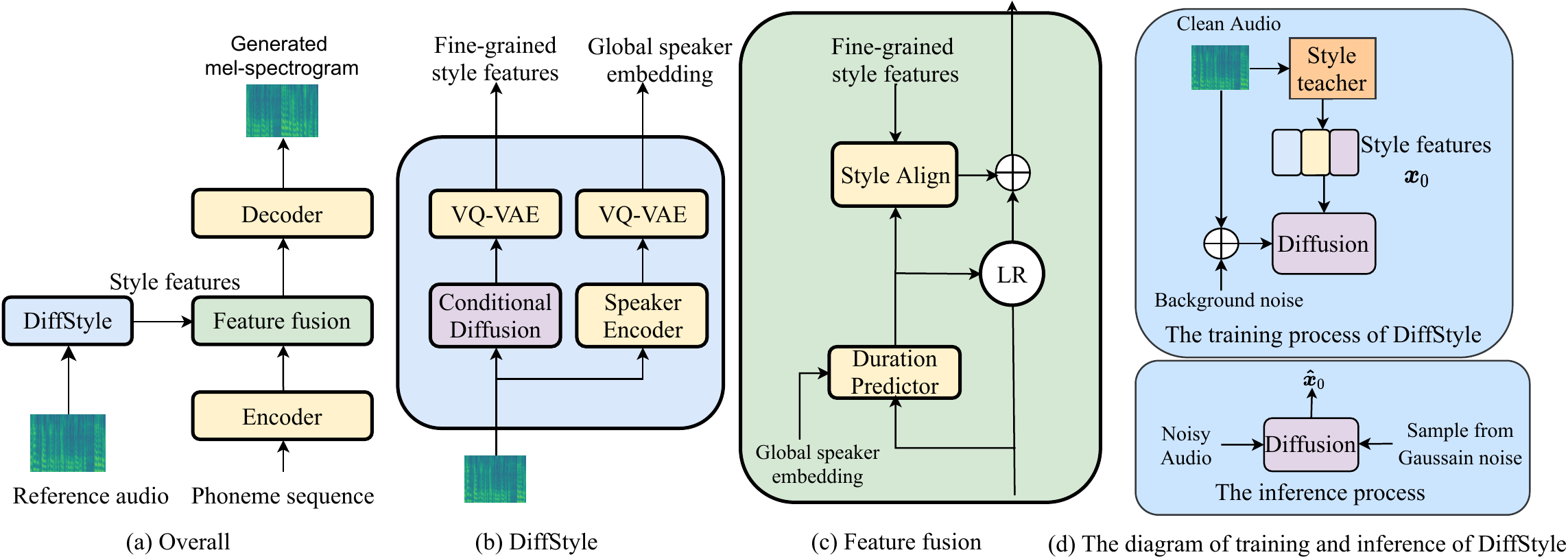}
  \caption{(a) shows the overview of our proposed NoreSpeech. (b) and (d) show the details of DiffStyle. In subfigure (c), LR denotes the length regulator, the decoder includes 4 Transformer blocks and one flow-based post-net, which is the same as \cite{huang2022generspeech}.}
  \label{fig:ml}
\end{figure*}

In this paper, instead of extracting style information from noisy reference audio, we propose to reconstruct style information by learning relevant parameters for distribution modeling. Specifically, we propose a knowledge distillation based conditional diffusion model \cite{ho2020denoising} that can directly generate deep style representation in a latent space conditioned on the noisy reference audio, named DiffStyle. 
Our DiffStyle is inspired by a popular diffusion-based speech enhancement model, CDiffuSE \cite{lu2022conditional}, which has shown that diffusion models are capable of restoring clean speech component from a noise-contaminated speech signal. The CDiffuSE model reconstructs clean waveform conditioned on a noisy spectrogram. The DiffStyle module in NoreSpeech is different from the CDiffuSE model in the following aspects: (1) Instead of generating highly stochastic time-domain audio signal (1 second composed of thousands of sample points), DiffStyle works on the more compressed frame-level features. (2) The CDiffuSE model reconstructs complete speech signal in time domain, while DiffStyle learns to generate prosody-related style features supervised by a pre-trained teacher model.
Furthermore, we explore two aspects of improving the generalization of style transfer: (1) To transfer style to a textual input from a length-mismatched reference utterance, we propose a parameter-free style-alignment module; (2) To transfer unseen speaking styles, we use a VQ-VAE module \cite{van2017neural} to map the style features into a controllable latent space, which has previously been shown to be an effective method \cite{huang2022generspeech}.



\section{Proposed method}
\label{sec:proposed method}
In this section, we first formulate the noise-robust expressive TTS model for style transfer. Then we overview our proposed \textbf{no}ise-\textbf{r}obust \textbf{e}xpressive TTS model (NoreSpeech), following which we introduce several critical components, including the style teacher model, DiffStyle, and feature fusion module.
\subsection{Problem formulation}
Style transfer aims to generate high-quality and similarity speech samples with previously unseen style (\textit{e.g.}, such as speaker identity and style) derived from a reference utterance. Thanks to the development in TTS, the state-of-the-art (SOTA) style transfer TTS models can realize good style transfer performance with high-quality reference audio. In this paper, we focus on a more challenging setting: the reference utterance contains background noise. Similar to Sytler \cite{lee2021styler}, we conjecture that speaker identity information can be extracted from noisy reference with a noise-robust speaker encoder. However, the style information will be affected by noise \cite{zhang2022hifidenoise}. Thus, the problem is to obtain style features from noisy reference similar to those from clean reference. 
\subsection{Overview} 
We adopt one of the SOTA style transfer TTS models, GenerSpeech \cite{huang2022generspeech} as the backbone.
The overall architecture of NoreSpeech has been showed in Fig. \ref{norespeech}. NoreSpeech is made up of four parts: (1) Encoder, which maps the phoneme sequence into deep representations; (2) DiffStyle, which generates style features based on a noisy spectrogram; (3) Feature fusion, which combines style and text features; (4) Decoder, which maps the features into mel-spectrogram. The encoder and decoder follow the same structure in the previous work \cite{huang2022generspeech}. 
\subsection{DiffStyle}
Fig. \ref{norespeech}(b) shows the diagram of DiffStyle, which includes three main parts: a conditional diffusion model, a speaker encoder, and two VQ-VAE \cite{van2017neural} blocks. The conditional diffusion model aims to generate fine-grained style features that represent the speaker's style, and the speaker encoder aims to generate a global speaker embedding that represents the speaker's identity. Both of them take the noisy reference utterance as input. In the following, we will introduce the speaker encoder and conditional diffusion models. 
\subsubsection{Speaker encoder} \label{speaker encoder}
We use a generalizable wav2vec 2.0 model \cite{baevski2020wav2vec} to capture the global speaker identity characteristics. Wav2vec 2.0 is a recently proposed self-supervised framework for speech representation learning. In our experiments, we add an average pooling layer and one fully-connected layer on top of the wav2vec 2.0 encoder, which allows us to finetune the encoder  on classification tasks. The AMsoftmax \cite{wang2018additive} loss is employed during finetuning.
\subsubsection{Conditional diffusion model} \label{con-diffusion}
Our target is training a conditional diffusion model that can generate noise-agnostic style features based on a noisy audio. To realize this, we adopt the idea of knowledge distillation, which uses a style teacher model to extract style features from clean speech, then the style features are used as the training objective of the diffusion model. 

\noindent
\textbf{Style teacher models} In this paper, we explore two types of style teacher: (1) Supervised learning (SL) based expressive TTS model, GenerSpeech \cite{huang2022generspeech}, which can effectively extract style features from clean speech. (2) Self-supervised learning (SSL) \cite{qian2020unsupervised, choi2021neural} based speech decomposition model, NANSY \cite{choi2021neural}, which can extract linguistic and style features from clean speech. We pre-train GenerSpeech and NANSY models in advance, then we take these two kinds of style teacher models to guide the training of NoreSpeech, denoting as NoreSpeech (T-SL) and NoreSpeech (T-SSL), respectively.

\noindent
\textbf{Diffusion model}
Diffusion probabilistic (diffusion for short) models \cite{sohl2015deep} have been proved as a powerful generation model in several important domains, \textit{e.g.} image \cite{dhariwal2021diffusion}, speech \cite{kong2020diffwave} and sound \cite{yang2022diffsound} fields. The basic idea of diffusion model is to train a neural network for reversing a diffusion process. Given i.i.d. samples $\{ \boldsymbol{x}_0 \in \mathbb{R}^D \}$ from an unknown data distribution $p_{data}(\boldsymbol{x}_0)$, diffusion models try to approximate $p_{data}(\boldsymbol{x}_0)$ by a marginal distribution $p_\theta(\boldsymbol{x}_0)= \int{p_\theta(\boldsymbol{x}_0, \cdots, \boldsymbol{x}_{T-1} | \boldsymbol{x}_T) \cdot p(\boldsymbol{x}_T)}dx_{1:T} $. \\
To implement our conditional diffusion model, we adopt the idea of conditional speech enhancement \cite{lu2022conditional}, which uses a shallow convolution layer $\tau_{\theta}()$ to reshape the noisy mel-spectrogram, then feeds it into a WaveNet-structure diffusion model. In our study, $\boldsymbol{x}_0$ represents style features rather than waveform. The training loss function can be defined as
\begin{equation}\label{vqvae loss}
\begin{aligned}
    \mathcal{L}_{\mathit{diff}} =  \mathbb{E}_{ST(\boldsymbol{y}_c),\boldsymbol{y}_n, \boldsymbol{ \epsilon} \sim \mathcal{N}(\boldsymbol{0},\boldsymbol{I}),t} \big{[} || \boldsymbol{\epsilon} - \boldsymbol{\epsilon}_{\theta}(\boldsymbol{x}_t,t,\tau_{\theta}(\boldsymbol{y}_n)) ||_2^2  \big{]} 
\end{aligned}
\end{equation}
where $ST$ denotes that style teacher model. $\boldsymbol{y}_c$ denotes clean mel-spectrogram, $\boldsymbol{y}_n$ denotes the noisy mel-spectrogram. $t$ is the index of time step. $\boldsymbol{\epsilon}_{\theta}$ denotes the learnable parameters. 
\subsubsection{Vector Quantization}
Considering the variability of generated style features, we use a Vector Quantization block \cite{van2017neural} to map the generated style features into a controllable latent space. We define a latent embedding space $\boldsymbol{e} \in \mathbb{R}^{K \times H} $ where $K$ is the size of the discrete latent space, and $H$ is the dimensionality of each latent embedding vector $\boldsymbol{e}_i$. In our experiments, we set $K=H=256$. To make sure that the representation sequence commits to an embedding and its output does not grow, a commitment loss is used:
\begin{equation}\label{vqvae loss}
    \mathcal{L}_{\mathit{c}} = || z_e(\boldsymbol{x}) -sg[\boldsymbol{e}] ||_2^2
\end{equation}
where $z_e(\boldsymbol{x})$ is the output of the vector quantization block, and $sg[\cdot]$ stands for the stop gradient operator.
\subsection{Feature fusion}
The feature fusion module aims to fuse the phoneme representation and style features. Considering the dimension mismatch between fine-grained style features and the output of the text encoder, we design a parameter-free style-align module to solve this problem. Assume that the time dimensions of style features and text features are $t_{style}$ and $t_{text}$, respectively. When $t_{style} \textless t_{text}$, we directly adopt a linear interpolation operation to upsample the style features. When $t_{style} \textgreater t_{text}$, we first calculate the ratio between $t_{style}$ and $t_{text}$, and then we average consecutive frames of style features based on the ratio to downsample the style features.
\begin{table}[t] \label{tab1}
\caption{Quality and style similarity results of style transfer.}
\label{tab:my-table0}
\begin{tabular}{c|cc}
\hline
\textbf{Method}       & \textbf{MOS}          & \textbf{SMOS}    \\ \hline
Reference             & 4.35 $\pm$ 0.09                  & -     \\
Reference(voc.)       & 4.32 $\pm$ 0.09                       & 4.31 $\pm$ 0.09 \\ \hline
FS2 (clean) \cite{ren2020fastspeech}          & 3.80 $\pm$ 0.09                  & 3.86  $\pm$ 0.09   \\
FS2 (nosiy)           & 3.73 $\pm$ 0.12                   & 3.72 $\pm$ 0.11  \\ \hline
Styler (noisy) \cite{lee2021styler}  & 3.86 $\pm$ 0.11                   & 3.89 $\pm$ 0.11 \\
GenerSpeech (clean) \cite{huang2022generspeech}  & 3.93 $\pm$ 0.11                   & 4.09 $\pm$ 0.11 \\
GenerSpeech (noisy)   & 3.87 $\pm$ 0.12                   & 3.81 $\pm$ 0.13 \\
GenerSpeech (denoise) & 3.89 $\pm$ 0.11                   & 3.95 $\pm$ 0.12  \\ \hline
NoreSpeech (T-SL) (noisy)         & 3.99 $\pm$ 0.10                    & 4.06 $\pm$ 0.11  \\
NoreSpeech (T-SSL) (noisy)        & \textbf{4.11 $\pm$ 0.09}      & \textbf{4.14 $\pm$ 0.09} \\ \hline
\end{tabular}
\end{table}
\subsection{Pre-training and loss function}
\textbf{Speaker encoder pre-training}
As section \ref{speaker encoder} described, we fine-tune the wav2vec 2.0 encoder on LibriTTS dataset, we implement this based on s3prl framework. \footnote{https://github.com/s3prl/s3prl}.

\noindent
\textbf{Pre-training style teacher} For GenerSpeech teacher, we reproduce GenerSpeech based on their paper \cite{huang2022generspeech}. The only difference is that we do not use emotion embedding. We train GenerSpeech on the LibriTTS dataset \cite{panayotov2015librispeech}. After that, we use the style adaptor of Generspeech to extract fine-grained prosodic features from clean speech. For NANSY teacher \cite{choi2021neural}, we first train NANSY \footnote{https://github.com/dhchoi99/NANSY} on LibriTTS dataset. Then, we use the pre-trained model to extract style features. 
\begin{table*}[t]
\caption{The AXY Preference test results. Preference is calculated based on 7-point score, where 0 is ``about the same''.}
\label{tab:my-table2}
\begin{tabular}{c|cccc|cccc}
\hline
\multirow{3}{*}{Baseline} & \multicolumn{4}{c|}{Parallel Style Transfer}                                                & \multicolumn{4}{c}{Non-Parallel Style Transfer}                                             \\ \cline{2-9} 
                          & \multicolumn{1}{c|}{\multirow{2}{*}{7-point score}} & \multicolumn{3}{c|}{Preference (\%)} & \multicolumn{1}{c|}{\multirow{2}{*}{7-point score}} & \multicolumn{3}{c}{Preference (\%)} \\
                          & \multicolumn{1}{c|}{}                               & Baseline    & Same    & NoreSpeech   & \multicolumn{1}{c|}{}                               & Baseline    & Same   & NoreSpeech \\ \hline
FS2                       & \multicolumn{1}{c|}{1.07}                           & 26\%        & 30\%    & 44\%         & \multicolumn{1}{c|}{1.48}                           & 35\%        & 17\%   & 48\%         \\
Styler                       & \multicolumn{1}{c|}{1.30}                           & 25\%        & 21\%    & 54\%         & \multicolumn{1}{c|}{1.26}                           & 29\%        & 21\%   & 50\%         \\
GenerSpeech               & \multicolumn{1}{c|}{1.20}                           & 29\%        & 27\%    & 44\%         & \multicolumn{1}{c|}{1.58}                           & 26\%        & 12\%    & 62\%         \\ \hline
\end{tabular}
\end{table*}

\noindent
\textbf{Loss function}
The final loss consists of the following parts 1) duration prediction loss $\mathcal{L}_{\mathit{dur}}$: MSE between the predicted and the ground-truth phoneme-level duration; 2) mel reconstruction loss $\mathcal{L}_{\mathit{mel}}$; 3) the negative log-likelihood of the post-net $\mathcal{L}_{\mathit{post}}$ \cite{huang2022generspeech}; 4) commitment loss $\mathcal{L}_{\mathit{c}}$: the objective to constrain vector quantization layer according to formula (2); 5) diffusion loss $\mathcal{L}_{\mathit{diff}}$ according to formula (1).
\section{Experiment}
\label{sec:experiments}
\subsection{Dataset, training setting and baseline models}
We train NoreSpeech on LibriTTS dataset \cite{panayotov2015librispeech}. To simulate noisy environments, we use the background sound from the acoustic scene classification task of DCASE 2019 Challenge \cite{mesaros2018multi}. All of the utterances of the noisy speech are mixed with noise sampled from DCASE with an SNR randomly chosen from 5 dB to 25 dB. To evaluate NoreSpeech, we randomly choose 20 sentences test data from LibriTTS test set, which does not appear on the training stage. We conduct preprocessing on the speech data: 1) converting the sampling rate of all data to 16kHz; 2) extracting the spectrogram with the FFT size of 1024, hop size of 256, and window size of 1024 samples; 3) converting it into a mel-spectrogram with 80 frequency bins. We train NoreSpeech for 200,000 steps. In the first 50000 steps, we directly feed the output of the style teacher to the feature fusion module. After that, we use the generated style features by the diffusion model as input. For the DiffStyle, the cosine schedule strategy $\beta_{t}=cos(0.5\pi \cdot \frac{(t/T+s)}{1+s})^2$ is used for any step $t$, where $s=0.008$ and $T=100$. We utilize HiFiGAN \cite{kong2020hifi} as the vocoder to synthesize waveforms from the generated mel-spectrogram. We conduct crowd-sourced human evaluations with MOS (mean opinion score) for naturalness and SMOS (similarity mean opinion score) \cite{min2021meta} for style similarity on Amazon Mechanical Turk. \\
\textbf{Baseline models} We compare the quality and similarity of generated audio samples of our NoreSpeech with other systems, including 1) Reference, the reference audio; 2) Reference (voc.), which means we convert the reference audio into mel-spectrograms and then convert them back to audio using HiFi-GAN; 3) FastSpeech 2 \cite{ren2020fastspeech}, which uses the speaker encoder to extract speaker embedding; 4) Styler \cite{lee2021styler}, which uses adversarial training and information bottleneck to eliminate noise;  5) GenerSpeech \cite{huang2022generspeech}, `clean', `nosiy', and `denoise' denote the types of reference audio.
\subsection{Experimental results}
Table 1 shows the MOS and SMOS comparisons between NoreSpeech and the baselines, and we have the following observations: (1) Noise has a significant impact on style transfer performance, \textit{e.g.} the SMOS of GenerSpeech drops from 4.09 to 3.81 when adding noise into reference audio. (2) GenerSpeech (denoise) denotes that we use one of the SOTA SE models \cite{ho2020denoising} to denoise the noisy reference, which can bring slight improvement. (3) Comparing to previous SOTA expressive TTS models (Styler and GenerSpeech), our NoreSpeech has better style transfer ability on noisy environment. (4) By comparing NoreSpeech (T-SL) and NoreSpeech (T-SSL), we can find that using an unsupervised speech decomposition (NANSY) as a teacher can bring better performance than using GenerSpeech as the teacher model. We conjecture that NANSY model can extract more robust style features from reference audio due to its self-supervised training strategy. We believe that better style teacher model can be explored to improve the performance of NoreSpeech. \\
To further evaluate NoreSpeech's style transfer ability, an AXY test \cite{huang2022generspeech} of style similarity is conducted to assess the style transfer performance, where raters are asked to rate a 7-point score (from -3 to 3) and choose the speech samples that sound closer to the target style in terms of style expression. We conduct parallel and non-parallel style transfer. \\
\textbf{Parallel style transfer (PST)} PST denotes that the text input is the same as the reference's content, Table 2 presents the results. Compared to FS2, Styler and Generspeech, our NoreSpeech has better style transfer performance. \\
\textbf{Non-parallel style transfer (N-PST)} 
We also explore the robustness of our NoreSpeech in N-PST, in which a TTS system synthesizes different text in the prosodic style of a reference signal. We can see that our NoreSpeech significantly improves the model to inform the speaking style, allowing a noisy reference sample to guide the robust stylistic synthesis of arbitrary text. This validates the effectiveness of the straight-forward text-style alignment module in NoreSpeech.
\section{Conclusions}
\label{sec:conclusion}
In this paper, we proposed a noise-robust expressive TTS model, named NoreSpeech. Benefitting from DiffStyle and style-align modules, NoreSpeech presents robust stylistic synthesis of arbitrary text, even if the reference audio includes noise. We proved that DiffStyle can be trained with two types of style teacher model, which shows DiffStyle can be further improved through training a better teacher model. We believe DiffStyle can also be used for other tasks (\textit{e.g.} image style transfer). In the future, we will explore better style teacher models and reduce the sample step in DiffStyle. 



\bibliographystyle{IEEE.bst}
\bibliography{refs.bib}

\end{document}